\documentstyle[epsf,graphicx]{mn}

\def\x2{$\chi^{2}$}

\def\x2{$\chi^{2}$}

\newcommand{\mincir}{\raise
-2.truept\hbox{\rlap{\hbox{$\sim$}}\raise5.truept\hbox{$<$}\ }}
\newcommand{\magcir}{\raise
-2.truept\hbox{\rlap{\hbox{$\sim$}}\raise5.truept\hbox{$>$}\ }}
\newcommand{\minmag}{\raise
-2.truept\hbox{\rlap{\hbox{$<$}}\raise6.truept\hbox{$<$}\ }}
\newbox\grsign \setbox\grsign=\hbox{$>$} \newdimen\grdimen 
\grdimen=\ht\grsign
\newbox\simlessbox \newbox\simgreatbox \newbox\simpropbox
\setbox\simgreatbox=\hbox{\raise.5ex\hbox{$>$}\llap
     {\lower.5ex\hbox{$\sim$}}}\ht1=\grdimen\dp1=0pt
\setbox\simlessbox=\hbox{\raise.5ex\hbox{$<$}\llap
     {\lower.5ex\hbox{$\sim$}}}\ht2=\grdimen\dp2=0pt
\setbox\simpropbox=\hbox{\raise.5ex\hbox{$\propto$}\llap
     {\lower.5ex\hbox{$\sim$}}}\ht2=\grdimen\dp2=0pt

\begin{document}

\title[The Apparent and Intrinsic Shape of the APM Galaxy Clusters] {The
Apparent and Intrinsic Shape of the APM Galaxy Clusters}

\author[S. Basilakos et al.]
       {S. Basilakos$^{1,2,3}$, M. Plionis$^{1}$, S. J. Maddox$^{4}$ \\
$^{1}$ National Observatory of Athens, I. Metaxa \& B. Pavlou, Lofos Koufou,
Palaia Penteli,
15236, Athens, Greece \\
$^{2}$ Physics Dept, University of Athens, Panepistimiopolis, 15783, Athens,
Greece \\
$^{3}$ Astrophysics Group, Imperial College London, Blackett Laboratory, 
Prince Consort Road, London SW7 2BZ, UK\\
$^{4}$ School of Physics and Astronomy, University of Nottingham,
       Nottingham NG7 2RD, UK
}

\maketitle


\begin{abstract}
We estimate the distribution of intrinsic shapes of APM
galaxy clusters from the distribution of their apparent shapes. 
We measure the projected cluster ellipticities using two alternative
methods.
The first method is based on moments of the discrete galaxy
distribution while the second is based on moments of the
smoothed galaxy distribution.
We study the performance of both methods using Monte Carlo cluster
simulations covering the range of APM cluster distances 
and including a random distribution of background galaxies.
We find that the first method suffers from severe systematic biases,
whereas the second is more reliable.
After excluding clusters dominated by substructure and quantifying the
systematic biases in our estimated shape parameters, we recover a
corrected distribution of projected ellipticities.
We use the non-parametric kernel method to estimate the smooth
apparent ellipticity distribution, and numerically invert a set of integral
equations to recover the corresponding distribution of intrinsic
ellipticities under the assumption that the clusters are either
oblate or prolate spheroids.
The prolate spheroidal model fits the APM cluster data best. 

\end{abstract}

\begin{keywords}
galaxies: clustering - Cosmology: Large Scale Structure in the Universe
\end{keywords}

\section{INTRODUCTION} 

Clusters of galaxies are the largest gravitationally collapsed objects
in the universe and their internal dynamics and morphologies provide
useful cosmological information.
In recent years many studies of cluster shapes and orientations
have showed that they are strongly elongated, maybe more so than
elliptical galaxies, and they tend to point towards their neighbours
(Carter \& Metcalfe 1980; Bingelli 1982; Di Fazio \& Flin 1988;
Plionis, Barrow \& Frenk 1991; De Theije, Katgert \& van Kampen 1995).
Plionis, Barrow and Frenk (1991) (hereafter PBF) have computed
ellipticities and major axis orientations for the largest up to date
sample of about 400 Abell clusters and found that their apparent
shapes are consistent with those expected from a population of prolate
spheroids.
Support to the prolate spheroidal case was presented recently by
Cooray (1999) analysing a sample of 25 Einstein X-ray clusters of Mohr
et al. (1995).
Struble and Ftaclas (1994) analysed a compilation of 344 Abell cluster
ellipticities and found that rich clusters are intrinsically more
spherical than poorer clusters.
In the same framework McMilan et. al (1989) studied the ellipticities
and orientations of 49 Abell clusters using Einstein X-ray data to 
trace the hot gas, and also found that the cluster potential is quite
flat although less so than that found in optical studies.
Buote \& Canizares (1996) analyzed ROSAT PSPC images for 4 Abell
clusters (including Coma) and, assuming hydrostatic
equilibrium,they found ellipticities of order 
$\epsilon_{mass}\simeq 0.40-0.55$
(see also Canizares \& Buote 1997).

Theoretical expectations regarding cluster shape and morphology have
been investigated via N-body simulations, which show that the intrinsic
shapes of simulated clusters are rather triaxial with an almost uniform
distribution of shapes between prolate and oblate spheroids (cf. Frenk
et al. 1988; Efstathiou et al. 1988). 
Detailed analysis of cluster morphological parameters and
substructure, utilising the concept of power ratios (cf. Boute \& Tsai
1994), can be used to constrain different cosmological models
(cf. Thomas et al 1998; Valdarnini, Ghizzardi \& Bonometto 1999).

It is obvious that information about the intrinsic shape of a cluster
is lost when projected  on the plane of the sky.  
Many different studies have attempted to recover the distribution of
intrinsic cluster shapes from the corresponding apparent distribution
using inversion techniques based on the assumption that their
orientations are random.

In this paper we use the APM catalogue (Dalton et al. 1997) to
measure the projected shape distribution, corrected for various
systematic effects, and hence attempt to estimate the intrinsic shape of
clusters. 
The APM clusters are typically as rich as Abell $R=0$ clusters, but due
to the careful identification procedure do not suffer from significant
projection effects.  

The plan of the paper is the following:
In Section 2 we describe the APM galaxy and cluster survey, in Section
3 we present our projected cluster shape determination method and by using 
Monte Carlo simulations we establish its statistical robustness.
We discuss how the foreground/background contamination (projection
effects) affects the projected cluster shapes and present a statistical
ellipticity correction procedure. 
The correction assumes that clusters are in dynamical equilibrium and
therefore we exclude clusters with strong substructure.
In Section 4 we invert the systematic bias-corrected projected
ellipticity distribution to recover the intrinsic one. 
Our conclusions are presented in Section 5.

\section{ THE APM DATA }

The APM survey covers an area of 4300 square degrees in the southern
sky ($b\leq -40^\circ$) and contains about 2.5 million galaxies brighter
than a magnitude limit of $b_{J}=20.5$.  
Details of the APM data can be found in Maddox et al. (1990a and
1990b), and Maddox, Efstathiou \& Sutherland (1996).
Here we present only a brief summary of the catalogue.
The survey was compiled from 185 survey plates from the UK Schmidt telescope
scanned by the Automatic Plate Measuring (APM) machine in Cambridge. 
The scanned region of each plate covers  $5.8^{\circ} \times
5.8^{\circ}$ of the sky, and since  neighbouring plate centers are
separated by $5^{\circ}$ this leads to $0.8^{\circ}$ overlaps along
plate boundaries.  
The data for each plate is stored separately to preserve the multiple
measurements in the overlap regions.
Extensive internal checks and external calibration have shown that the
plate-to-plate zero-point error has an rms of 0.06 magnitudes, and
that large-scale photometric gradients are even smaller. 
The IRAS and COBE all-sky maps show that the galactic obscuration in
this region of the sky is typically 0.06 magnitudes introducing
comparable uncertainty in the photometry.  
The image profiles and shapes were used to classify them into 
galaxies, stars  and blended stars. Visual checks and deeper CCD
images show that the classification leads to galaxy samples which are
90-95\% complete with contamination of 5-10\% from non-galaxies. 

Dalton et al (1997) applied an object cluster finding algorithm to the
APM galaxy data, and so produced a list of galaxy clusters, most of
which have subsequently been spectroscopically confirmed as clusters. 
The cluster finding algorithm consists of two main steps: 
The first step  uses a percolation algorithm to link all pairs of
galaxies with separations $< 0.7$ the mean inter-galaxy separation. 
All mutually linked pairs are joined together to form groups, and the
groups with more than 20 galaxies are identified as  candidate clusters.
In the second step, an iterative routine is applied to each candidate
to estimate  the richness and characteristic apparent magnitude of
galaxies within a search radius of  $0.75 \; h^{-1}$ Mpc. 
This produced a list of 957 clusters with $z_{est} \mincir 0.1$ and APM
richness of more than 40 galaxies, corresponding roughly to Abell
richness class 0.   
The angular diameter of the search radius is set to be consistent with
the distance estimated from the apparent magnitude of galaxies in the
search radius.

For our present analysis we cross-correlated the cluster positions
with the APM galaxy survey, and for each cluster selected all galaxies
falling within a distance of $1.2 \; h^{-1}$ Mpc from the cluster
center. 
Since this is a larger radius than used in the cluster identification,
some clusters near to the survey boundaries do not have complete data
over the full circle.
We found that 54 of the APM clusters are affected, and have simply
rejected them from the sample, leaving 903 clusters which we use
in our analysis.

\section{Projected Cluster Shapes}

\subsection{Basic Methods}

In order to estimate the APM cluster shapes we use the moments of
inertia method (cf. Carter \& Metcalfe 1980; Plionis, Barrow \& Frenk
1991)
The galaxy equatorial positions are transformed into an equal area
coordinate system, centered on the cluster center, using:
$x =(Ra_{g}-Ra_{cl}) \times \cos(\delta_{cl})$ and
$y =\delta_{g}-\delta_{cl}$, where subscripts $g$ and $cl$ refer to
galaxies and the cluster, respectively. 
We then evaluate the moments:
$I_{11}=\sum\ w_{i}(r_{i}^{2}-x_{i}^{2})$,
$I_{22}=\sum\ w_{i}(r_{i}^{2}-y_{i}^{2})$,
$I_{12}=I_{21}=-\sum\ w_{i}x_{i}y_{i}$, with $w_{i}$ the
statistical weight of each point. 
Note that because the inertia tensor is symmetric we have
$I_{12}=I_{21}$.
Diagonalizing the inertia tensor
\begin{equation}\label{eq:diag}
{\rm det}(I_{ij}-\lambda^{2}M_{2})=0 \;\;\;\;\; {\rm (M_{2} \;is \; 
2 \times 2 \; unit \; matrix.) }
\end{equation}
we obtain the eigenvalues $\lambda_{1}$, $\lambda_{2}$, from which we
define the ellipticity of the configuration under study by:
$\epsilon=1-\lambda_{2}/\lambda_{1}$, with $\lambda_{1}>\lambda_{2}$.
The corresponding eigenvectors provide the direction of the principal axis.

This basic shape estimation method is applied using two alternative
methods:

\noindent (1) Discrete case ($w=1$): In which we use the individual
galaxies to determine the cluster shape.
At first all galaxies within a small radius from the cluster center
($\sim 0.1 \; h^{-1}$ Mpc) are used to define the initial value of the
cluster shape parameters, then the next nearest galaxy is added
consecutively to the initial group and the shape is recalculated until
we include all galaxies within a limiting radius of our choice (usually
$\sim 0.75 \; h^{-1}$ Mpc).

\noindent
(2) Smooth case ($w=\delta$): In which we use the smoothed galaxy
distribution on a grid $N_{gr} \times N_{gr}$, where the surface density
of the $j^{th}$ cell is:
\begin{equation}\label{eq:smoo}
\rho_{j}(x_{gr})=\frac{\sum_{i}\rho_{j}(x_{i})F(x_{i}-x_{gr})}
{\int F(x_{i}-x_{gr}) {\rm d}^{2}x} \;\;\;\;\;\; .
\end{equation}
The sum is over the distribution of galaxies at positions $x_{i}$,
with the Gaussian kernel being:
\begin{equation}\label{eq:gaus}
F(x_{i}-x_{gr})=\frac{1}{2\pi R_{sm}}
\exp\left(-\frac{(x_{i}-x_{gr})^{2}}{2R_{sm}^2} \right)
\end{equation}
All cells having a density above a chosen threshold are used to define
the cluster shape, with cell-weights corresponding to the density fluctuation:
\begin{equation}
w_j = \frac{\rho_{j}(x_{gr})-\langle \rho \rangle }{\langle \rho
\rangle}
\end{equation}
where $\langle \rho \rangle$ is the mean projected APM galaxy density.
Note that the apparent magnitude limit means that the APM clusters at
large distances contain only the few most luminous cluster galaxies.
Therefore the smoothing radius $R_{sm}$ must increase with cluster
distance to obtain a continuous density field free of discreteness
effects.
We used sets of Monte Carlo cluster simulations with the same richness
and depth distributions as the APM clusters to chose $R_{sm}$, as a
function of distance, so that it minimises discreteness effects and
optimises the performance of our shape measuring algorithm. 
We find a roughly linear relation, fitted by: $R_{sm}\simeq
2.1\times 10^{-4}r+0.023$, where $r$ is the cluster distance.

To estimate the cluster shape parameters we use cells above a
threshold, defined as the mean $\langle \delta \rangle$ value
of all grid-cells within a chosen radius (measured in $h^{-1}$ Mpc), 
$R_{\delta}$, from the cluster center.
After testing a range of values on mock clusters with the
characteristics of the APM catalogue, we concluded that a value of
$R_{\delta} \simeq 0.36$ $h^{-1}$ Mpc is a good compromise: 
too small a radius leads to inadequate sampling of the cluster region
within $\sim 1$ $h^{-1}$ Mpc; 
too large a radius extends the analysis to the very end of the sampled 
cluster region and thus artificially sphericalizes the clusters while it
also increases the contamination effects by including relatively more
background galaxies.
Note that the highest density peak does not necessarily coincide with 
the listed APM cluster center. We have therefore measured 
moments about the position of
the highest density peak, found within 0.5 $h^{-1}$ Mpc of the nominal
centre. In only a few cases is there any significant difference
between these two cluster centers. 
In any case, using either does not change the
results of our analysis, although on an individual basis it can have a
marked difference to the derived cluster ellipticity, especially
if the two center distance is large.

In figure 1 we present the frequency distribution of the maximum
cluster radius; ie. the distance between the cluster center and the
most distant grid-cell that is used in the shape determination
procedure, for two values of $R_{\delta}$.
It is evident that there is a range of radii samples, depending
slightly also on the value of $R_{\delta}$, the most common of which
$\sim 0.65$ $h^{-1}$ Mpc.
\begin{figure}
\mbox{\epsfxsize=8.7cm \epsffile{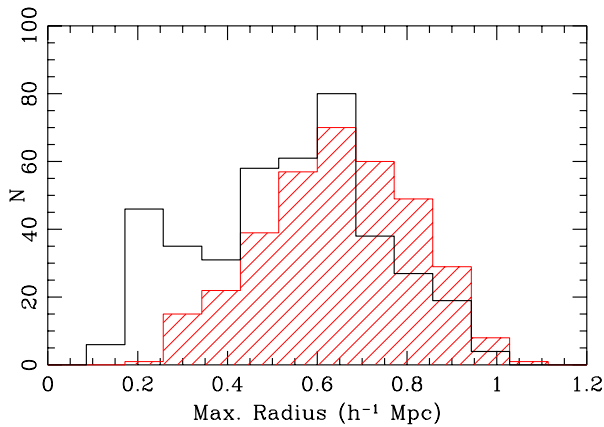}}
\caption{Distribution of maximum APM cluster radius
used to define the cluster shape parameters for $R_{\delta}=0.24$
(thick line) and $R_{\delta}=0.36$ (hatched). 
Only the 405 APM clusters with no significant substructure
are used (see section 3.4)}
\end{figure}

\subsection{Testing the Performance of our Method}
The two procedures to determine cluster shape parameters may be
biased by the unavoidable presence of foreground and background
galaxies projected on the clusters, and may also have systematic
biases due to the methods themselves.
In this section we investigate the robustness of the two methods
in recovering the true projected cluster shape in the presence of
discreteness effects and the galaxy background. 

To this end we generate a large number of mock clusters, resembling in
appearance dynamically relaxed structures; ie., having no substructure
and King-like surface brightness profiles:
\begin{equation}\label{eq:PROF}
\Sigma(r)\propto \left[ 1+\left(\frac{r}{r_{c}}\right)^{2} \right]
^{-\alpha} \;\;, 
\end{equation}
where $r_{c}$ is the cluster core radius and $\alpha$ is the slope
parameter.
Different values of the slope and core radius have been found in
different studies, spanning the range $0.6 \mincir \alpha \mincir 1$
and $0.1 \mincir r_c \mincir 0.25$ $h^{-1}$ Mpc (cf. Bachall \& Lubin
1993; Girardi et al. 1995; Girardi et al. 1998).
The latter study, using the ENACS optical sample of Abell clusters,
found a median value of $\alpha\simeq 0.65_{-0.07}^{+0.05}$.
We create our mock clusters by randomly generating galaxies
having $\alpha = 0.65$, $r_{c}=0.18$ and input ellipticities
of our choice. We have tested that small variations of these
parameters do not alter significantly our results.

The expected global background at each cluster  can be
estimated by:
\begin{equation}\label{eq:PROF1}
N_{back}={\rm d}\Omega_{i}\int_{0}^{z_{max}} \langle \rho(z) \rangle z^{2} dz
\end{equation}
where $z_{max}$ is the maximum redshift of galaxies in the APM
catalogue ($\sim 0.3$), ${\rm d}\Omega_{i}$ is the solid angle covered
by the cluster, given by ${\rm d}\Omega_{i}=2\pi (1-cos\theta_{i})$ for a
cluster with angular radius $\theta_{i}$, and
$\langle \rho(z) \rangle$ is the mean APM galaxy density 
at redshift $z$, obtained by integrating the APM galaxy luminosity function,
$\Phi(M,z)$, allowing for evolution (Maddox, Efstathiou \& Sutherland
1996):
\begin{equation}\label{eq:mean}
\langle \rho(z) \rangle=\int^{M_{max}}_{M_{min}(z)} \Phi(M,z) dM \;,
\end{equation}
with $M_{min}(z)$ the minimum absolute magnitude that a galaxy at a
redshift $z$ can have and still be included in the APM catalogue,
limited in apparent magnitude by 
$m_{lim}=20.5$, ie; $M_{min}(z)=m_{lim}-42.38-5\log z+5\log h -3z \;\;.$
The points in figure 2 show the number of APM galaxies counted within a
radius of 1.2 $h^{-1}$ Mpc from each cluster center, and the line
shows the  predicted global background, $N_{back}$. 
Both are decreasing functions of distance due to the apparent magnitude
limit of the APM galaxy catalogue.
\begin{figure}
\mbox{\epsfxsize=8.5cm  \epsffile{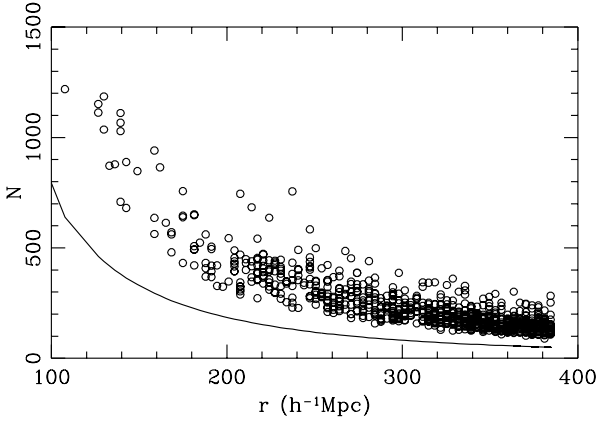}}
\caption{The APM cluster richness (number of galaxies within a
$1.2 \; h^{-1}$ Mpc radius) and the expected number of background galaxies
as a function of distance.}
\end{figure}

For our Monte-Carlo simulations we generate a random galaxy background, with
number density given from eq.(\ref{eq:PROF1}), in a circular area of radius
equal to the semi-major axis of the ellipse. 
As an example we plot in figure 3 the smooth galaxy density distribution
of two mock clusters with $\epsilon_{th}= 0.5$ at distances
$r=200 \;h^{-1}$ Mpc and $380 \;h^{-1}$ Mpc,
respectively and having the typical APM cluster richness at that distance.
As expected the random background tends to sphericalize the clusters.

The question that we want to answer now is: ``Given an input cluster
ellipticity what is the most probable measured ellipticity recovered
by our methods?"
We present a case study of the mock cluster at a distance of $200
\;h^{-1}$ Mpc, having a range of input projected ellipticities of
width 0.1.
We generate 100 Monte-Carlo realizations of the cluster for each input
ellipticity. In the left panel of figure 4 we present the measured 
mean ellipticities, and their scatter for both methods in the absence
of a background galaxy distribution.
Both methods give similar results, with the first method recovering
(for $\epsilon_{th} \magcir 0.1$) exactly the input (correct)
ellipticities.
\begin{figure}
\mbox{\epsfxsize=9cm \epsffile{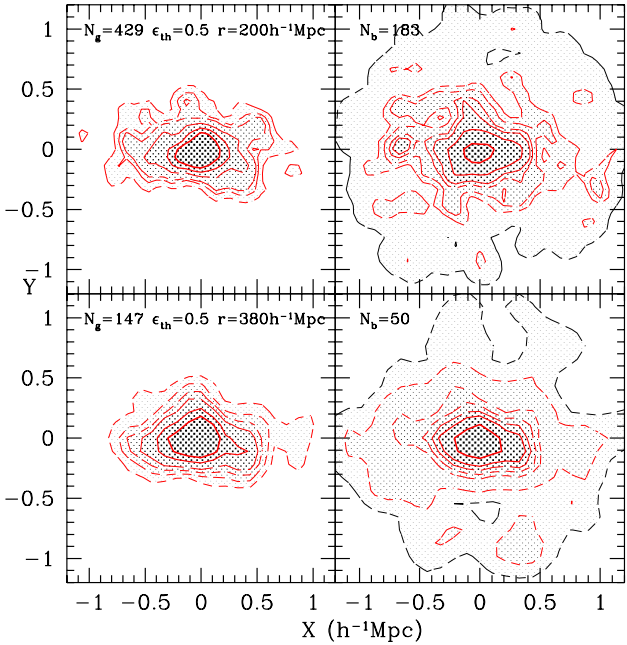}}
\caption{The density contours for two mock clusters using a King-like
profile and input ellipticity $\epsilon_{th}=0.5$. }
\end{figure}

\begin{figure}

\mbox{\hspace{-0.4cm} \epsfxsize=9cm \epsffile{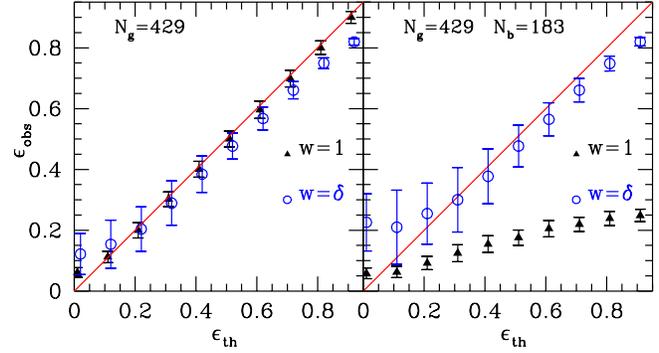}}
\caption{Performance of our shape determination method for a simulated
cluster at $200 \;h^{-1}$ Mpc containing 429 galaxies, by comparing the
input and measured ellipticities, using either $w=1$ or $w=\delta$ weights.
The left panel presents the case for no
background galaxy contamination. The right panel
presents the more realistic situation in which we superimpose a random
background galaxy distribution according to eq.(\ref{eq:PROF1}).}
\end{figure}
The right panel of figure 4 presents the results for the same mock cluster
but now we include the random background galaxy distribution, appropriate
for the distance of the cluster. The first method breaks down and severely
underestimates the input ellipticities for $\epsilon >0.1$.
It is evident that only the $w=\delta$ method performs relatively well
in the presence of the galaxy background and from now on we will be
using only this method.

Performing many tests with mock clusters at different
distances, we conclude that the grid method has a variable performance
as a function of distance, tending to increasingly underestimate the 
ellipticity of elongated clusters as a function of distance
(by about 0.1-0.15). 
For clusters with $\epsilon_{th} \mincir 0.2-0.25$ the relation
between input and recovered ellipticity is non-monotonic, which means
it is not possible to correct the measured ellipticities of APM
clusters for the systematic biases which are evident in the right panel
of figure 4. 
These numerical tests have served mostly to choose between the two
shape determination procedures, rather than to derive a robust
ellipticity correction procedure.

\subsection{Correcting systematic ellipticity biases}

Since cluster ellipticities cannot be corrected on an individual basis
we must apply  a statistical correction to the ellipticity distribution to
deal with all the above mentioned
systematic effects. 
In order to do this, we need to answer a slightly different question
from the one we posed previously. The relevant question is:
``What is the distribution of input (correct) ellipticities from which a
measured ellipticity can be obtained?"
For each APM cluster with measured ellipticity, $\epsilon_{obs}$, we
generate 50 mock Monte-Carlo clusters for each ellipticity bin of
width 0.05, spanning the whole range $\epsilon_{th}\in (0,1)$.
These mock clusters, 1000 in total, 
have the same number of cluster and background
galaxies and are placed at the same distance as the one observed.
We measure the ellipticity of the mock clusters, $\epsilon_{mock}$, 
and derive its distribution function per
each input ($\epsilon_{th}$) ellipticity bin. 
We then measure, assuming Gaussian statistics, how many standard deviations
the original APM cluster measured ellipticity ($\epsilon_{obs}$)
differs from the derived mean mock value, 
$\langle \epsilon_{mock}\rangle$, and assign to the corresponding input 
ellipticity ($\epsilon_{th}$) bin the resulting probability.
Therefore, for each input ($\epsilon_{th}$) ellipticity bin $i$ 
we estimate the 
probability of being the correct ellipticity from which the APM
measured one could have resulted; ie,
\begin{equation}
P^{i}(\epsilon_{th})=1 - P(z_{i})
\end{equation}
where $z_i=|\epsilon_{obs}-\langle
\epsilon^{i}_{mock}\rangle|/\sigma_i$.
If for example, for some bin $i$, the value of $|\epsilon_{obs}-\langle
\epsilon^{i}_{mock}\rangle| = 4\sigma_i$ then $z_i=4$, $P(z_i)\simeq 1$
and thus the probability of $\epsilon^{i}_{th}$ being the correct APM cluster
ellipticity is $P^{i}\simeq 0$.
Doing so for all bins we derive the full probability distribution
function of input (correct) ellipticities, $P(\epsilon_{th})$, from
which our measured APM cluster ellipticity could have resulted.
For each APM cluster we have therefore generated in total 1000 mock 
clusters (50 per $\epsilon_{th}$ bin) which provide an estimate 
$P(\epsilon_{th})$ for that cluster.
The final step in estimating the corrected cluster ellipticity
distribution is to add the contribution from each cluster by randomly
sampling its $P(\epsilon_{th})$ $N$ times (we have used $N=20$). 
We stress that the correction procedure is applied separately to each
individual APM cluster which means that the corrections take into
account the variations in performance of our shape determination
method as a function of different cluster richness and distance.

As an illustrative example we plot in figure 5 the $P(\epsilon_{th})$
distribution for three distant clusters ($r\simeq 320$ $h^{-1}$ Mpc)
with $\epsilon_{obs} \simeq 0.2, 0.5$ and 0.7 respectively.
For the $\epsilon_{obs}=0.2$ case the $P(\epsilon_{th})$ is quite flat
in the range $0\mincir \epsilon_{th} \mincir 0.2$, having a long tail up to
$\epsilon_{th} \simeq 0.6$. 
The $\epsilon_{obs}=0.5$ case has a $P(\epsilon_{th})$ distribution
that peaks at $\sim 0.65$ but also has a significant contribution from
lower ellipticities. Finally for the $\epsilon_{obs}=0.7$ case the most 
probable value of $\epsilon_{th}$ is $\sim 0.9$.
These facts serve to show that in order to produce the correct
apparent cluster ellipticity distribution it is essential to use a
statistical procedure and sample each $P(\epsilon_{th})$ distribution
adequately.
\begin{figure}
\mbox{\epsfxsize=8cm \epsffile{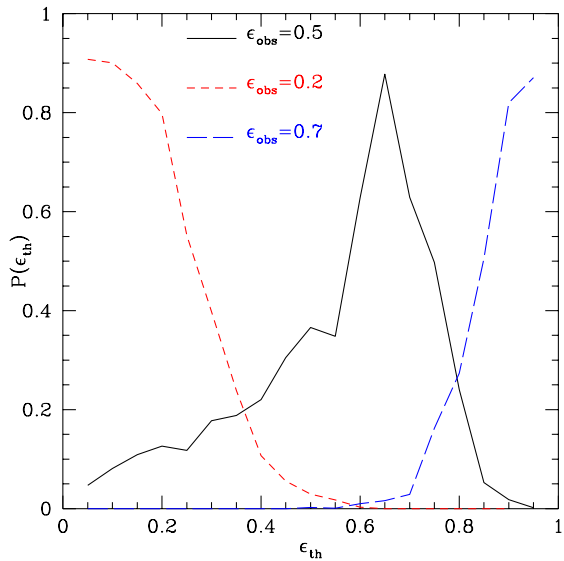}}
\caption{The probability functions for some characteristic measured
ellipticities ($\epsilon_{obs}=0.2, 0.5$ and 0.7) of three APM clusters
containing $\simeq 140$ galaxies and at a distance of $\sim 320$
$h^{-1}$ Mpc. } 
\end{figure}

\subsection{Substructure of APM clusters}

The correction procedure of the ellipticity distribution is based on
the assumption that the clusters under study have a smooth King-like
density profile.
Therefore we need to exclude from our sample clusters that
exhibit evidence of significant substructure.
The number of clusters with substructure, expected to be undergoing
dynamical evolution, is an unsettled issue but of great importance
since it provides information of the mean density of the universe
(cf. Richstone et al. 1992; Dutta 1995; Buote 1998; Thomas et al. 1998).
In an $\Omega=1$ universe, clusters continue to form even today and 
therefore one expects more substructure than in a
low $\Omega$ universe.
Identifying real sub-clumps in clusters is a difficult problem in
general, since one has to work either in two-dimensions, in which
projection effects can significantly affect the visual structure of
clusters, or in redshift space, where distortions due to knowledge of
only the radial velocity component can again distort the true pattern.
Several studies indicate that at least $\sim 30\% - 50\%$ (cf. West
1994 and references therein; Jones \& Forman 1999) of rich clusters
have strong substructure in their gas or galaxy distribution within
$\simeq 1 \; h^{-1}$ Mpc of the cluster center.

Here, we present the main steps of our substructure identification
method, the full details and results will be presented in a
forthcoming paper.
We work on the smoothed density field, as described in section 3.1.
For each overdensity threshold, estimated for $R_{\delta}=0.12, 0.24$
and 0.36 $h^{-1}$ Mpc, we select all grid-cells with overdensities
above the specific threshold.
We then connect all cells that have common borders to
create multiple clumps.
In all cases we accept only clumps that are within
$\simeq 0.75 \;h^{-1}$ Mpc of the highest cluster peak; we have found
this scale  to be optimal for reliable substructure identification
by validating our methods and results using ROSAT data for a subsample 
of 22 clusters (Kolokotronis et al. 2000), which also corresponds 
to the counting radius used in the APM cluster finding algorithm 
(cf. Dalton et al. 1997).

Investigation of the number and size of these clumps as a function of
overdensity threshold, provides the following categorisation:
\begin{itemize}
\item No substructure (69 clusters): Clusters with one clump in all
overdensity levels. 
\item Weak substructure (336 clusters): Multiple clumps only at the
lowest overdensity level or at the highest two overdensity levels but
where the second in size clump is $ <20 \%$ of the total cluster size
(cf. Richstone et al. 1992). 
\item Strong substructure (498 clusters): Multiple clumps where the
second in size clump is $ \magcir 20 \%$ of the total cluster size.
\end{itemize}
We have investigated the robustness of our substructure characterisation
procedure, as a function of different $R_{\delta}$, and found only
small variations, consisting mainly of a movement of APM clusters
between the ``no" and ``weak" substructure categories.
We have also verified that due to the random galaxy background,
coupled with discreteness effects, it is common to find ``weak"
substructure even in our mock clusters which by construction have no
substructure.
Therefore we exclude from our shape determination analysis only those
APM clusters that were found to have ``strong" substructure.

\section{True Cluster Shapes}

In order to find the intrinsic ellipticity distribution assuming that
clusters are all oblate or prolate spheroids, we use a standard method
based on the kernel estimator.

\subsection{Kernel estimator}

General reviews of kernel estimators are given by Silverman (1986) and
Scott (1992) but the applications to the astronomical data are given
by Vio et al. 1994, Tremblay \& Merritt (1995) and Ryden (1996).
Here we review the basic steps of the Kernel method, following the
notation of Ryden (1996). 
For each estimated ellipticity $\epsilon$, we estimate the axial ratio, 
$q = \left( \frac{1-\epsilon}{1+\epsilon} \right) ^{1/2}$. 
Given the sample of axis ratios $q_{1},q_{2},....,q_{N}$ for $N$
clusters, the kernel estimate of the frequency distribution is defined
as:
\begin{equation}\label{eq:ker1}
\hat{f}(q)=\frac{1}{Nh} \sum_{i}^{N}\ K\left(\frac{q-q_{i}}{h}\right) \; \; ,
\end{equation}
where $K(t)$ is the kernel function, defined so that 
\begin{equation}\label{eq:ker2}
\int_{-\infty}^{+\infty}\ K(t) {\rm d}t=1 \; \; ,
\end{equation}
and $h$ is the ``kernel width" which determines the balance between
smoothing and noise in the estimated distribution.
In general the value of $h$ is chosen so that the expected value of the
integrated mean square error between the true, $f(q)$, and estimated, 
$\hat f(q)$, distributions, 
$\int_{-\infty}^{+\infty} \left[\hat{f}_{K}(x)-f(x) \right]^{2} {\rm d}x$,
is minimised (cf. Vio et al. 1994; Tremblay \& Merritt 1995).
In this work we estimate the $h$ using a very common approach presented
by Silverman (1986), Vio et al. (1994) and Ryden (1996) in which:
\begin{equation}\label{eq:width}
h=0.9 A N^{-1/5}
\end{equation}
where $N$ is the number of the clusters and $A=\min(\sigma, Q_4/1.34)$,
with $Q_4$ the interquartile range.
There are three common choices for the kernel function $K(t)$ which
have quadratic, quartic and Gaussian forms (cf. Tremblay \& Merritt
1995).
Many studies have shown that the choice of a kernel function does not
in general affect the estimates, and they differ trivially in their
asymptotic efficiencies.
We have chosen a Gaussian kernel:
\begin{equation}\label{eq:gaus2}
K(t)=\frac{1}{\sqrt{2\pi}} e^{-t^{2}/2} \; \; .
\end{equation}

In order to obtain physically acceptable results with $\hat{f}(q)=0$
for $q<0$ and $q>1$, we apply reflective boundary conditions which
means that the Gaussian kernel is replaced with:
\begin{equation}\label{eq:Ker3}
K(q,q_{i},h)=K\!\!\left(\frac{q-q_{i}}{h}\right)+
K\!\!\left(\frac{q+q_{i}}{h}\right)+K\!\!\left(\frac{2-q-q_{i}}{h}\right)
\end{equation}
This also ensures the correct normalization, $\int_{0}^{1} \hat{f}(q) 
{\rm d}q =1$. 
For a discussion of reflective boundary conditions see Silverman
(1986) and Ryden (1996). 
\begin{figure}
\mbox{\epsfxsize=8cm \epsffile{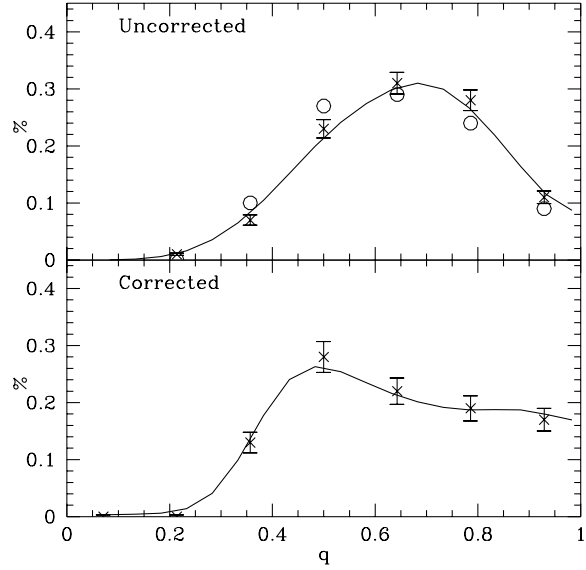}}
\caption{
In the top panel the crosses show the uncorrected distribution of
apparent axial ratios for the complete sample of 903 APM clusters 
with the solid line being the smooth distribution from the nonparametric 
kernel estimator.  The 
open circles correspond to the distribution of the 405 APM clusters
with no significant substructure.
In the lower panel we present the corrected axial ratio distribution 
for the later clusters.}
\end{figure}
The crosses in figure 6 show the projected axial ratio distributions
with the Poisson 1$\sigma$ error bars and the solid lines show the
kernel estimate $\hat{f}$ with width $h=0.075$.
In the top panel we present our results for the uncorrected
distribution of all 903 APM clusters; which can crudely be fitted by a
Gaussian with $\langle q \rangle \simeq 0.65$ and standard deviation
$\simeq 0.15$. 
In the bottom panel we present the corrected distribution using
the 405 APM clusters free of significant subclustering.
It is obvious that the two distributions are significantly different,
with the peak of the corrected distribution having moved to lower $q$'s but
with an extended contribution of apparently quasi-spherical objects.

\subsection{Inversion method}
The relation between the apparent and intrinsic axial ratios, is
described by a set of integral equations first investigated by Hubble
(1926).
These are based on the assumptions that the orientations are random
with respect to the line of sight, and that the intrinsic shapes can
be approximated by either oblate or prolate spheroids.
There is no physical justification for the restriction to oblate or
prolate but it greatly simplifies the inversion problem.
Furthermore, if the intrinsic shape of clusters is triaxial or a mixture of
the two spheroidal populations then there is no unique inversion (PBF).
Writing the intrinsic axial ratios as $\beta$ and the estimated distribution
function as $\hat N_o(\beta)$ for oblate spheroids, and $\hat
N_p(\beta)$ for prolate spheroids then the corresponding distribution
of apparent axial ratios is given for the oblate case by:
\begin{equation}\label{eq:apaobl}
\hat{f}(q)=q\int_{0}^{q}\frac{\hat{N}_{\circ}(\beta) {\rm d}\beta}
{(1-q^{2})^{1/2}(q^{2}-\beta^{2})^{1/2}}
\end{equation}
and for the prolate case by:
\begin{equation}\label{eq:apaprol}
\hat{f}(q)=\frac{1}{q^{2}}\int_{0}^{q}\frac{\beta^{2}\hat{N}_{p}(\beta) 
{\rm d}\beta}
{(1-q^{2})^{1/2}(q^{2}-\beta^{2})^{1/2}} \; \; .
\end{equation}
Inverting equations (eq.\ref{eq:apaobl}) and (eq.\ref{eq:apaprol})
gives us the distribution of real axial ratios as a function of the measured
distribution:
\begin{equation}\label{eq:oblate}
\hat{N}_{o}(\beta)=\frac{2\beta (1-\beta^{2})^{1/2}}{\pi} \int_{0}^{\beta}
\ \frac{\rm d}{{\rm d}q}\left(\frac{\hat{f}}{q} \right)\frac{{\rm d}q}
{(\beta^{2}-q^{2})^{1/2}}
\end{equation}
and
\begin{equation}\label{eq:prolate}
\hat{N}_{p}(\beta)
=\frac{2(1-\beta^{2})^{1/2}}{\pi\beta} \int_{0}^{\beta}
\ \frac{\rm d}{{\rm d}q}(q^{2}\hat{f})
\frac{{\rm d}q}{(\beta^{2}-q^{2})^{1/2}} \; \; .
\end{equation}
with $\hat{f}(0)=0$. In order for $\hat{N}_{p}(\beta)$ and $\hat{N}_{o}(\beta)$
to be physically meaningful they should be positive for all
$\beta$'s. 
Following Ryden (1996), we numerically integrate eq.(\ref{eq:oblate})
and eq.(\ref{eq:prolate}) allowing $\hat{N}_{p}(\beta)$ and
$\hat{N}_{o}(\beta)$ to take any value. 
If the inverted distribution of axial ratios has significantly
negative values, a fact which is unphysical, then this can be  
viewed as a strong indication 
that the particular spheroidal model is unacceptable.

In figure 7 we present the uncorrected and corrected intrinsic axial
ratio distributions.
The uncorrected distribution for both spheroidal models takes negative
values; over the range $\beta \magcir 0.7$ for the oblate case, and
$\beta \magcir 0.8 $ for the prolate case.
Using the corrected apparent axial-ratio distributions $\hat{f}(q)$,
the oblate model produces negative values of $N$ for $\beta \magcir
0.5$ and $\beta \mincir 0.2 $, but the prolate one provides a distribution
of intrinsic axial ratios that is positive over the whole $\beta$
range.

\begin{figure}
\mbox{\epsfxsize=9cm \epsffile{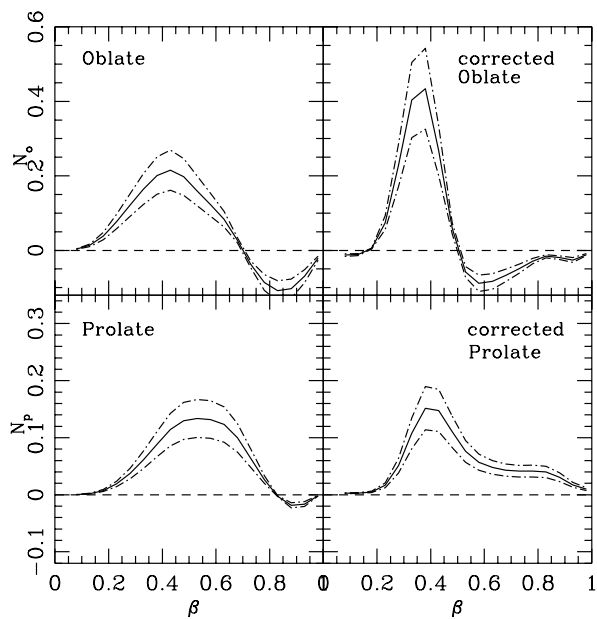}}
\caption{The distribution of intrinsic APM cluster axial ratios and 
its $1 \sigma$ range, for the corrected and uncorrected samples, 
assuming that clusters are either oblate or prolate spheroids.}
\end{figure}

This suggests that the APM cluster shapes are better represented by
that of prolate spheroids rather than oblate, which is in agreement
with PBF and Cooray (1999).
However, it is probably not realistic to assume a population of pure oblate
or prolate spheroids but rather of triaxial ellipsoids, in which case
the inversion procedure is not unique (see PBF).
Nevertheless, our results strongly suggest that cluster prolateness 
should be a dominant feature.

\section{Conclusions}

We have measured the projected ellipticities of all APM clusters using 
moments either of the individual
galaxy distribution or of the smoothed galaxy distribution above
some overdensity threshold.
We have performed large sets of Monte Carlo simulations in order to
test the statistical robustness of the two procedures, and conclude
that the first method is strongly affected by the presence of
background galaxies whereas the second method better recovers the
underline true cluster projected ellipticity.

We devised a statistical Monte-Carlo procedure to correct the distribution of
cluster ellipticities for the systematic errors introduced by
discreteness effects, background galaxies and the method itself.
The procedure involves estimating first the probability distribution
of the true projected ellipticity for each APM cluster, and then
random sampling this probability distribution to estimate the
cluster's contribution to the corrected overall projected cluster
ellipticity distribution.
This method works well only for clusters that appear relaxed, with no
significant substructure.
`Therefore we have excluded, from our final cluster sample which contains
405 clusters, all the APM
clusters that show evidence of significant substructure.
Prior to the exclusion of these clusters the uncorrected axial ratio
distribution of the whole APM cluster sample can be crudely
approximated by a Gaussian with a mean of $\simeq 0.65$ and a standard
deviation of $\simeq 0.15$. 
The corrected apparent axial-ratio distribution is significantly
different showing a bump at $q\simeq 0.46$ and having a significant
contribution from apparently quasi-spherical systems.

Using the nonparametric kernel procedure we obtain a smooth estimate
of the apparent APM cluster axial-ratio distribution.
We assume that the APM clusters are a homogeneous population of either
oblate or prolate spheroids and numerically invert the apparent
distribution to obtain the intrinsic distribution.  The most
acceptable model is provided by that of prolate spheroids. This result
supports the view by which clusters form by accretion of smaller units
along the large-scale structure (filament) in which they are embedded
(cf. West 1994; West, Jones \& Forman 1995). 
Such an accretion process would happen preferentially
along the cluster major axis, which is typically aligned with the
nearest cluster neighbour (cf. Bingelli 1982; Plionis 1994 and
references therein).

Since cluster shapes and substructure are sensitive cosmological probes 
(Evrard et al. 1993), we plan to
investigate these issues further using APM clusters and compare our
results with theoretical expectations, provided by high-resolution
N-body simulations of different cosmological models (cf. Thomas et
al.1998). 

\section* {Acknowledgements}
S.B. thanks the British council and Greek State Fellowship Foundation
for financial support (Contract No 2669).
We thank Dr. E. Kolokotronis for fruitful discussions,
Dr. M.Kontizas for her constant support and Dr. D.Buote, the referee, 
for his positive comments.

\end{document}